# A Novel Ferroelectric Rashba Semiconductor

Gauthier Krizman*, Tetiana Zakusylo, Lakshmi Sajeev, Mahdi Hajlaoui, Takuya Takashiro, Marcin Rosmus, Natalia Olszowska, Jacek J. Kołodziej, Günther Bauer, Ondrej Caha, and Gunther Springholz*

G. Krizman, T. Zakusylo, M. Hajlaoui, T. Takashiro, G. Bauer, G. Springholz
Institut für Halbleiter und Festkörperphysik
Johannes Kepler Universität
Altenberger Strasse 69, 4040 Linz, Austria
E-Mail: gauthier.krizman@jku.at, gunther.springholz@jku.at

L. Sajeev, O. Caha
Department of Condensed Matter Physics
Masaryk University
Kotlárská 2, 61137 Brno, Czech Republic

M. Rosmus, N. Olaszowska, J.J. Kołodziej
National Synchrotron Radiation Centre SOLARIS
Jagiellonian University
Czerwone Maki 98, 30-392 Krakow, Poland

J. J. Kołodziej
Faculty of Physics, Astronomy and Applied Computer Science
Jagiellonian University
Ul. Prof. Stanislawa Lojasiewizca 11, 30-348 Krakow, Poland

**Abstract**: Fast, reversible, and low-power manipulation of the spin texture is crucial for next generation spintronic devices like non-volatile bipolar memories, switchable spin current injectors or spin field effect transistors. Ferroelectric Rashba semiconductors (FERSC) are the ideal class of materials for the realization of such devices. Their ferroelectric character enables an electronic control of the Rashba-type spin texture by means of the reversible and switchable polarization. Yet, only very few materials have been established to belong to this class of multifunctional materials. Here, $Pb_{1-x}Ge_xTe$ is unraveled as a novel FERSC system down to nanoscale. The ferroelectric phase transition and concomitant lattice distortion is demonstrated by temperature dependent X-ray diffraction, and its effect on electronic properties are measured by angle-resolved photoemission spectroscopy. In few nanometer-thick epitaxial heterostructures, a large Rashba spin-splitting is exhibiting a wide tuning range as a function of temperature and Ge content. Our work defines $Pb_{1-x}Ge_xTe$ as a high-potential FERSC system for spintronic applications.

# 1. Introduction

Ferroelectric Rashba semiconductors (FERSC) have been recently disclosed as a new class multifunctional material to enrich electronic and spintronic device technologies[1–9]. The unique feature of FERSC is the fundamental breaking of the inversion symmetry caused by a ferroelectric (FE) lattice distortion, which leads to a large spin splitting of the electronic band structure in k-space by the Rashba effect[10,11] (Fig. 1(a)). The direction of the spin polarization, i.e., the helicity of the spin texture is locked to the FE polarization. This means that in a FERSC the spin polarization can be externally controlled and reversed by an applied electric field via a non-volatile and switchable poling process[12,13]. This remarkable property is singular to this class of multifunctional materials and is sought-after for spintronic applications such as spin field effect transistors, non-volatile and bipolar memories as well as programmable transistors for nematics and logic operations[9,13–16].

The development of FERSC demands materials exhibiting ferroelectricity, semiconductor properties and a sizeable Rashba at the same time. Recent theoretical studies have suggested a number of potential FERSC candidates like complex oxides[17–19], perovskites[20–22] or 2D materials[15,23,24], but so far, FERSC have been demonstrated experimentally only for the IV-VI class of semiconductors (see Fig. 1(b)). The key representative is $\alpha$-GeTe[3,4,12,13,25–29], which is FE below its Curie temperature of $T_c \sim 700$ K[30] and displays a giant Rashba effect[12,25,27–29,31,32]. This is due to the large rhombohedral lattice distortion in which the cation $Ge^{2+}$ and anion $Te^{2-}$ sublattices[30] are shifted with respect to each other by as much as 0.3 Å along the <111> direction[33], as illustrated by Fig. 1(a). This induces a permanent electric dipole that accounts for a macroscopic FE polarization and the concomitant Rashba effect on the electronic band structure. Moreover, by means of an applied electric field, a switching of the spin polarization by controlling the FE polarization has been demonstrated. Although a giant Rashba effect has been also observed for the highly polar compounds BiTeI[34,35] and BiTeBr[36], because these are non-ferroelectric, the spin polarization in them cannot by switched and reversed permanently by an electric field.

The drawback of $\alpha$-GeTe for FERSC applications is its intrinsic high p-type conductivity that arises from the high density of electrically active Ge vacancies in the crystal lattice. This results in a rather large intrinsic hole concentration above $10^{20}$ cm$^{-3}$ that cannot be compensated or controlled by counter doping[4,37], which impedes efficient control of the FE polarization due to leakage currents. In addition, $\alpha$-GeTe has an indirect band gap[38], which makes it not suitable for optical devices. Last but not least, the bulk FE Rashba effect of $\alpha$-GeTe is superimposed and partially screened by the giant Rashba effect of the localized surface states[27,39] caused by the tellurium surface termination favored by the free surface energy of the system. Thus, for $\alpha$-GeTe the experimental identification of the bulk FE Rashba effect requires detailed analysis of photoemission spectroscopy data by ab initio density functional theory calculations to sort out between the intrinsic bulk and the extrinsic surface Rashba effects. It is noted that SnTe (Fig. 1(b)) is another potential candidate for FERSC, but it suffers from the same problems as $\alpha$-GeTe (except for the indirect gap) and exhibits a relatively low $T_c \sim 100$ K[40] and thus, does not provide a solution for device applications[13,31,41].

**In this work,** we pursue an alternative approach to overcome the limitations of $\alpha$-GeTe, based on the conversion of paraelectric (PE) PbTe by GeTe doping to a FERSC with superior properties for device applications. PbTe is a versatile representative of the IV-VI compounds due to its direct band gap, as well as the three orders of magnitude lower carrier concentration as compared to GeTe and SnTe. Moreover, it features very high carrier mobilities exceeding $10^6$ cm$^2$/V.s at low temperatures[42] and can be effectively doped *p*- as well as *n*-type[43]. Due to the more ionic character compared to GeTe, PbTe crystallizes in the centrosymmetric PE cubic rock salt structure. However, it is very close to a PE-FE phase transition (see Fig. 1(b)), which is signified by the pronounced phonon softening[44,45] and strong increase of the dielectric constant[46–48] at low temperatures that follows a Curie behavior and yields by extrapolation a negative $T_c \approx -70$ K[49] (see Fig. 1(c)). As a result, already a minute doping of PbTe with GeTe immediately converts $Pb_{1-x}Ge_xTe$ into a ferroelectric material with a critical temperature that rises super linearly with Ge concentration[50], reaching a $T_c$ at room temperature already at $x_{Ge}$ = 13%. At the same time, the direct band gap, low carrier concentration and high mobility is retained[51–53].

Here we develop molecular beam epitaxy (MBE) for the growth of high quality $Pb_{1-x}Ge_xTe$ films and quantum confined ferroelectric heterostructures and study their structural and electronic properties by combining x-ray diffraction (XRD) and high-resolution angle-resolved photoemission spectroscopy (ARPES). Based on unprecedented ARPES data, we reveal that $Pb_{1-x}Ge_xTe$ unites all key features of a FERSC, namely, that (i) the FE can be tuned over a wide range by Ge doping, that (ii) $Pb_{1-x}Ge_xTe$ displays a giant Rashba effect that is absent for the PE phase and that its appearance is precisely correlated with the onset of the FE phase transition, (iii) the Rashba spin texture persists for ultra-thin films, which is essential for an integration into microelectronic devices, that (iv) the behavior of the Rashba effect precisely follows the Landau-Ginzburg theory of a 2$^{nd}$ order phase transition, and that finally (v) the magnitude of the Rashba splitting and the FE polarization are *linearly* related. All taken together, we demonstrate intricate interplay between the structural and electronic properties of FERSC, and thus, establish for the first time a complete picture of the unique properties of this novel class of multifunctional materials that provide great potentials for device applications.

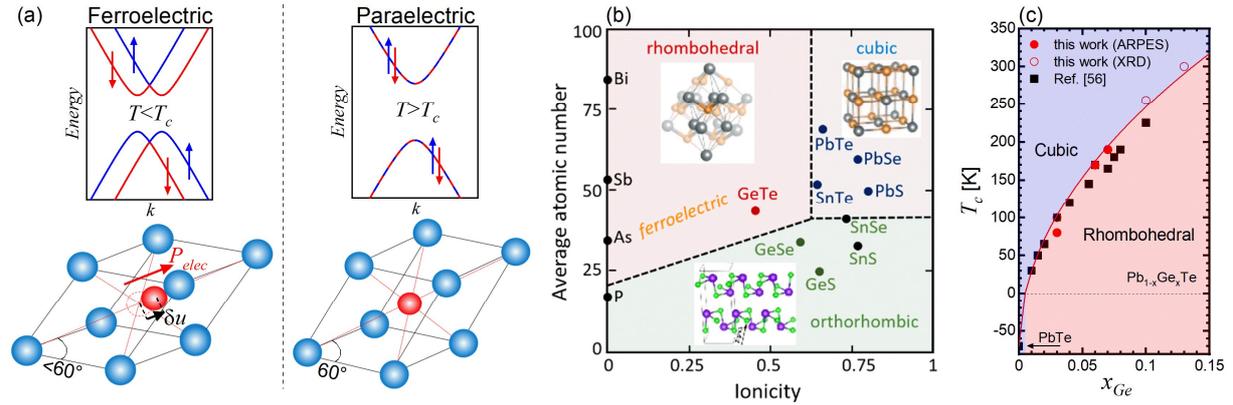

**Figure 1. IV-VI and $Pb_{1-x}Ge_xTe$ systems as FERSC. (a)** Comparison of the electronic band structure of a rhombohedral distorted FE semiconductor such as GeTe (left) with that of a PE cubic semiconductor such as PbTe (right). As shown in the lower panel, in the rhombohedral phase, the cations (red) are shifted by $\delta u$ with respect to the anions (blue) along the <111> direction. This induces an electric dipole responsible for a Rashba-type spin splitting of the bands illustrated in the top panels. In the cubic phase, shown on the right, the centrosymmetric cation site is located at the center of the anionic sublattice and thus, the electric dipole is zero and the bands are Kramer's spin degenerate. **(b)** Polymorphism in IV-VI class of compounds illustrating the formation of different structural phases and ferroelectricity as a function of ionicity and average number of electrons per atom (after Ref.[54]). **(c)** Curie temperature of $Pb_{1-x}Ge_xTe$, showing the FE phase transition as a function of temperature. The solid line is a fit to our experimental data, giving $T_c(x) = 1000\sqrt{x} - 70$ [K].

## 2. Results and discussion
### 2.1. Growth

Pb$_{1-x}$Ge$_x$Te films and quantum well (QW) heterostructures were grown by MBE on (111) BaF$_2$ substrates[55–57] using PbTe and GeTe as source materials (see methods section). In this way, perfect two-dimensional (2D) layers were achieved as shown by Fig. 2. A key feature of Pb$_{1-x}$Ge$_x$Te growth is the very strong temperature dependence of the Ge incorporation into the epilayers. This is because the vapor pressure of GeTe is three orders of magnitude higher than that of PbTe[58,59]. As a result, at the common IV-VI MBE growth temperatures > 350°C[55], the re-evaporation rate of GeTe from the surface is so high (> 5 Å/sec) that no GeTe is actually incorporated in the PbTe epilayers.

To overcome this limitation, the growth temperature for Pb$_{1-x}$Ge$_x$Te has to be drastically reduced below 300°C to suppress reevaporation and achieve a sizeable Ge concentration. This is demonstrated by Fig. 2(d) that shows the (222) XRD spectra of a series of Pb$_{1-x}$Ge$_x$Te epilayers grown on PbTe buffer layers at different temperatures from 250 to 300°C as indicated. With decreasing growth temperature and a fixed PbTe to GeTe flux ratio of 10:1, one sees that the Pb$_{1-x}$Ge$_x$Te layer peak strongly shifts away from the PbTe peak, which appears at the same diffraction angle independent of the growth temperature. This signifies that $x_{Ge}$ increases from 5 to 10% just by decreasing the growth temperature to 250 °C if one considers the change of the Pb$_{1-x}$Ge$_x$Te lattice parameter according to the Vegard's law as:

$$a_{PbGeTe}(x_{Ge}) = 6.462 - 0.472\, x_{Ge} \quad [\text{Å}]$$

valid for the cubic phase at room temperature[60,61].

For the Pb$_{1-x}$Ge$_x$Te films on PbTe buffer layers shown in Fig. 2, however, the lattice of the layer is strained to the that of the buffer layer, which means that the out-of-plane lattice parameter is expanded due to the Poisson ratio and thus, the change in the perpendicular direction is amplified by factor of two to $\Delta a_z(x_{Ge}) = 0.999\text{Å}\, x_{Ge}$. This is because for $x_{Ge} < 0.15$, the thickness of the 50 nm films is below the critical thickness for strain relaxation. This is confirmed by Fig. 2(b), which shows that the spacing of the (02) streaks of the reflection high energy electron diffraction (RHEED) patterns, recorded during the Pb$_{1-x}$Ge$_x$Te/ PbTe growth, does not change with layer thickness and remains equal to that of the PbTe buffer layer. The perfect 2D growth and the formation of atomically flat surfaces is further witnessed by the pronounced RHEED intensity oscillations shown in Fig. 2(b,c) that persist to more than 40 monolayers. This is corroborated by the atomic force microscopy (AFM) image shown in Fig. 2(e) evidencing that the surface consists of flat terraces separated only by few single monoatomic steps. The well-developed finite thickness Laue fringes around the XRD peaks in Fig. 2(d) also demonstrate the high quality of the PbGeTe/PbEuTe interfaces. By low temperature growth, we have achieved Pb$_{1-x}$Ge$_x$Te layers with $x_{Ge}$ up to 0.13, which is well above the Ge solubility limit of ≈0.06 at 300 °C previously determined for bulk material[61].

Different layer structures were grown and investigated in this work. Apart from thick films (from 1 to 5 μm) used for the assessment of the FE phase transition by temperature-dependent XRD measurements, Pb$_{1-x}$Ge$_x$Te heterostructures with QW thickness down to 8 nm were fabricated by growth on wide band gap Pb$_{0.9}$Eu$_{0.1}$Te barriers (100 nm) pre-deposited on μm-thick fully relaxed PbTe buffer, as shown by the inset of Fig. 2(d).

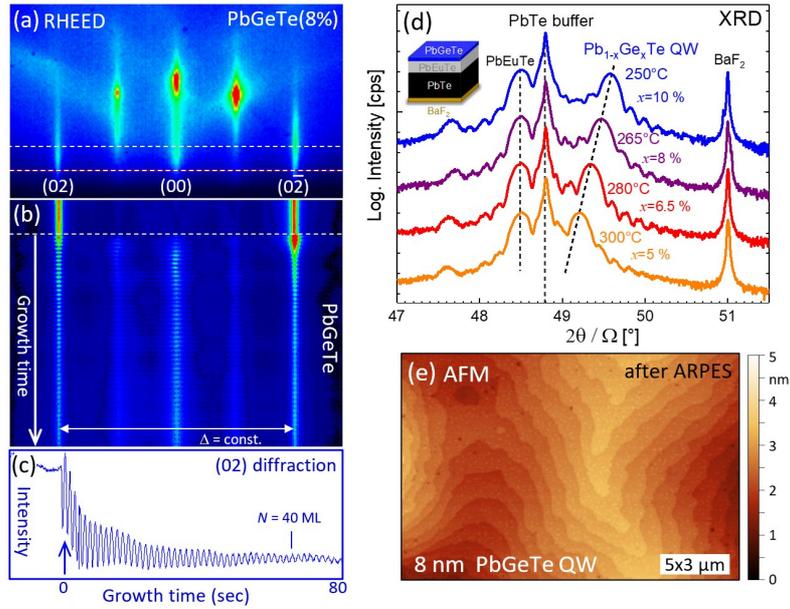

**Figure 2. MBE growth and characterization of $Pb_{1-x}Ge_xTe$ quantum well heterostructures. (a)** RHEED patterns and **(b)** evolution of the intensity profile along the dashed lines in (a) measured in situ during MBE growth of $Pb_{0.92}Ge_{0.08}Te$ on PbTe (111) at a temperature of 280°C. **(c)** Perfect pseudomorphic 2D growth is signified by the pronounced RHEED intensity oscillations and the constant ∆-spacing between the diffraction peaks. **(d)** Radial (222) XRD spectra of four 50 nm $Pb_{1-x}Ge_xTe$ QW layers grown at different temperatures from 250° to 300°C at the same GeTe/PbTe flux ratio of 1:10. As indicated by the dashed lines, the $Pb_{1-x}Ge_xTe$ layer peaks indicates a strong increase of the Ge concentration $x_{Ge}$ when the growth temperature is decreased. The finite thickness Laue fringes indicate the high quality of the QW interface. The sample structure is shown as insert. **(e)** AFM surface image of a $Pb_{0.93}Ge_{0.07}Te$ QW obtained after ARPES investigations, revealing a flat surface with only single monoatomic steps and a root mean square roughness below 0.5 nm.

Concerning the electrical properties, Ge-doping slightly increases the tendency of cation vacancy formation that leads to a slight p-doping of the layers. This latter is, however, orders of magnitude weaker that for GeTe, and thus, this effect can be easily compensated by doping with Bi atoms during growth[43]. Bi acts as a donor and leads to an n-doping. Therefore, $Pb_{1-x}Ge_xTe$ layers can be made p- or n- type depending on the extrinsic Bi-doping. Accordingly, in our thick samples and our heterostructures, carrier densities from $10^{17}$ to $10^{19}$ $cm^{-3}$ n- or p-type were obtained. The carrier mobility is found as high as 1 000 $cm^2/V.s$ for low Ge content samples ($x_{Ge} \lesssim 4$ %) at room temperature, and decreases to several hundred of $cm^2/V.s$ for higher Ge content. As a result, $Pb_{1-x}Ge_xTe$ displays superior transport properties[52,53].

## 2.2. Ferroelectric phase transitions and sublattice displacement

To reveal the FE phase transition, temperature dependent XRD measurements were performed on a series $Pb_{1-x}Ge_xTe$ films with $0 < x_{Ge} < 0.13$. The results are presented in Fig. 3, where in panels (a,b) radial XRD scans over the (333) and (444) Bragg reflections for $x_{Ge}$= 0.06 are depicted. Evidently, with decreasing temperature, the $Pb_{0.96}Ge_{0.06}Te$ diffraction peak is seen to shift parallel to the $BaF_2$ substrate due to the shrinking of the lattice parameter by the ordinary thermal contraction of the materials, however, at a critical temperature of $T_c$ = 160 K the peaks shift reverses its direction, signifying the onset of the cubic to rhombohedral phase transition that leads to an elongation of the unit cell in the [111] direction (see Fig. 1(a)). In addition, below $T_c$ the $Pb_{0.96}Ge_{0.06}Te$ peak splits up into two peaks because the rhombohedral distortion can occur in any of the four equivalent <111> directions of the crystal lattice. This means that FE domains with different elongation directions are formed, where for the "p"-domains the elongation is along the [111] surface normal, whereas for the "o"- domains the elongation is along one of the other three $\langle\bar{1}11\rangle$ distortions. For this reason, their out-of-plane lattice parameter is smaller than for the "p"-domains and this difference leads to the splitting observed in the experiments. This is corroborated by the (333) reciprocal space maps (RSMs) displayed in Fig. 3(c,d) that show that above $T_c$ only a single Pb$_{1-x}$Ge$_x$Te diffraction peak appears whereas a below $T_c$ two well-separated peaks are observed, one for the "p" domain and one for the "o" domains. Accordingly, this splitting is another indication for the structural phase transition

As shown by Fig. 3(b), the splitting is strongly temperature dependent because the rhombohedral distortion below $T_c$ increases with decreasing temperature. This is also signified by Fig. 3(e), where the measured out-of-plane lattice parameter $a_{111}$ is plotted as a function of temperature for the Pb$_{1-x}$Ge$_x$Te samples with different compositions. As indicated by the arrows, in all cases the slope of $a_{111}(T)$ changes sign clearly at a certain critical temperature, which marks the onset of the cubic/rhombohedral phase transition, with the $T_c$ increasing from 100 to 160 and 260 K for $x_{Ge}$ = 0.03, 0.06 and 0.1, respectively. As shown by Fig. 1(c) this dependence can be described by the relation $T_c(x_{Ge}) = 1000\sqrt{x_{Ge}} - 70$ [K].

It is noted, however, that the rhombohedral distortion alone is not sufficient evidence for a FE state because for a FE polarization to merge, the lattice centrosymmetry must be broken by a shift of the anion/cation sublattices with respect to each other, as shown schematically in Fig. 1(a). To assess this shift, we evaluate the intensity evolution of the (*hkl*) Bragg peaks that are governed by the change of the structure factor $|F_{hkl}|$ occurring when atom positions are shifted within the unit cell. This effect is particularly pronounced for *odd* (*hkl*) reflections where the waves scattered by the anion and cation planes are exactly out of phase in the centrosymmetric $Fm\bar{3}m$ cubic structure. This means that even minute changes in the anion/cation lattice plane distance due to the transition to the non-centrosymmetric $R3m$ phase yields a relatively large change in diffraction intensity, whereas for the even (*hkl*) the change is negligible because the scattered waves are in phase. This is exactly what we observe for the Pb$_{1-x}$Ge$_x$Te (333) and (444) Bragg peaks as shown in Fig. 3(a,b). Below $T_c$, the (333) intensity rises rapidly by more than a factor of two, whereas the (444) is essentially constant (see supplemental information for details).

Because the sublattice shift $\delta u$ in Pb$_{1-x}$Ge$_x$Te is along the [111] direction, $\delta u$ can be directly determined from the odd (*hkl*) intensity ratio between the FE and PE phases according to[40]

$$\frac{I_{hkl}^{FE}}{I_{hkl}^{PE}} = cos^2\{2\pi(h+k+l)\delta\} + \left(\frac{f_A+f_C}{f_A-f_C}\right)^2 sin^2\{2\pi(h+k+l)\delta\}$$

where $f_A$ and $f_C$ are the anion and cation form factors and $\delta$ denotes the normalized deviation of the lattice planes from the center position by $\delta u = 2a\sqrt{3}\delta$. From our measurements, we thus quantitatively obtain the dependence of $\delta u$ as a function of $x_{Ge}$ and temperature presented in Fig. 3(f), which directly evidences that the onset of the PE-FE phase transition is perfectly correlated with the appearance of a sublattice shift at exactly the same critical temperature. Below $T_c$, evidently $\delta u$ increases and approaches a saturation value that linearly increases with increasing Ge content, reaching a value of $\delta u \approx 0.11$ Å for $x_{Ge}$ = 0.1 at T=72 K. This corresponds to a change in the anion/cation lattice plane spacing as large as 3%, that is about half of the GeTe value[30,33].

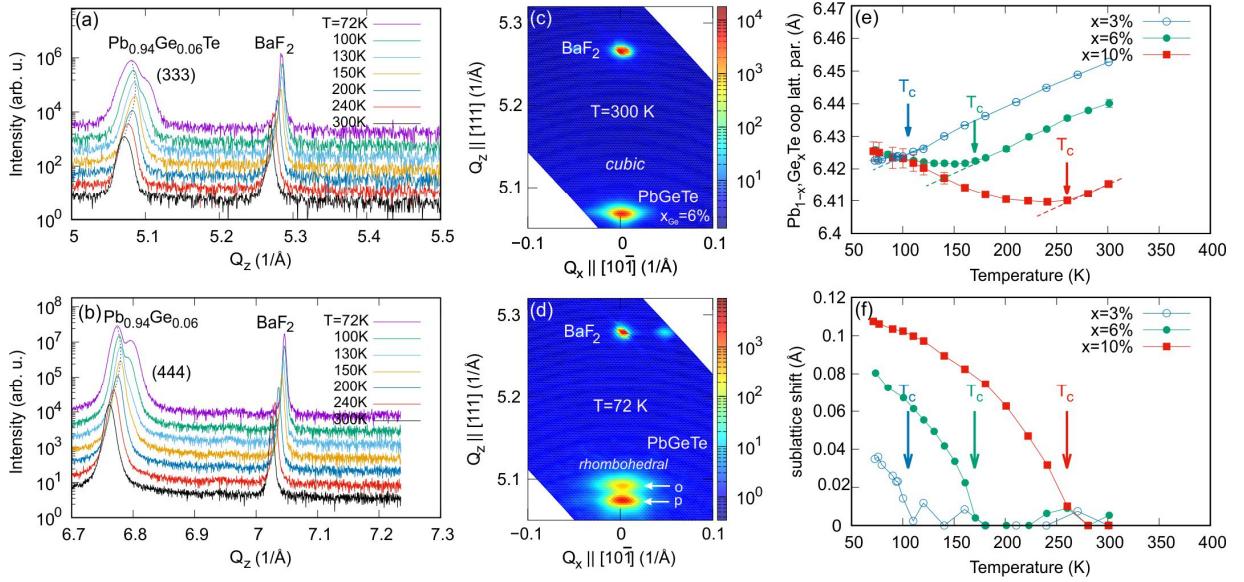

**Figure 3. Ferroelectric structural phase transition in Pb$_{1-x}$Ge$_x$Te. (a,b)** XRD scans across the (333) and (444) Bragg reflections measured at various temperatures for Pb$_{0.94}$Ge$_{0.06}$Te on BaF$_2$ (111). **(c,d)** Reciprocal space maps around the (333) reciprocal lattice point at 300 and 72 K, respectively. **(e)** Temperature dependence of the out-of-plane lattice parameter of Pb$_{1-x}$Ge$_x$Te with 0<x<0.1, showing the elongation of the unit cell along the [111] direction occurring below the critical temperature $T_c$. **(f)** Same for the measured anion/cation sublattice shift $\delta u$ along the [111] direction determined from the change of the intensity ratio between (333) and (444) diffraction peaks. The critical temperatures are indicated by the arrows.

## 2.3. Ferroelectric Rashba effect

ARPES was employed to resolve the impact of ferroelectricity on the electronic band structure. To this end we have prepared Pb$_{1-x}$Ge$_x$Te heterostructures (9 nm-thick QWs) to obtain single domain films in which the formation of "o"-domains is completely suppressed. These films were grown on 100 nm thick wide band gap barrier layers of Pb$_{0.9}$Eu$_{0.1}$Te in order to effectively confine the electrons and holes in the FE Pb$_{1-x}$Ge$_x$Te layer, which, as shown in Fig. 4, strongly enhance the signature of the Rashba effect. ARPES measurements were performed around the high symmetry points $\bar{\Gamma}$ and $\bar{M}$ of the 2D Brillouin zone (BZ) where the band extrema are located, as represented in Fig. 4(a).

ARPES measurements of PbTe and Pb$_{0.93}$Ge$_{0.07}$Te QWs around the $\bar{\Gamma}$ and $\bar{M}$-points are shown in Fig. 4(b-e) together with their fit using $\boldsymbol{k}\cdot\boldsymbol{p}$ theory as detailed below for cubic and rhombohedral lattices (see the Methods section). Due to the quantization of the electronic states in the QW, a large number of quantum confined states show sharp dispersions (< 20 meV line width) both for the PbTe as well as the Pb$_{0.93}$Ge$_{0.07}$Te QW, which are perfectly reproduced by the $\boldsymbol{k}\cdot\boldsymbol{p}$ calculations using the parameters given in the supplementary material. Evidently, the quantized subbands of Pb$_{0.93}$Ge$_{0.07}$Te QW appear to be split in the $k_\parallel$-direction, which is a clear indication for the FE Rashba effect. This phenomenon is obviously absent in the PE case illustrated by PbTe (Fig. 4(b,c)).

For the $\bar{\Gamma}$-point, the energy level spacing between the different quantum confined states is found much smaller than for the $\bar{M}$-points. This is due to the difference between the electronic dispersions of the two types of valleys in the quantization direction as a result of the huge anisotropy in this system (see the schematic ellipsoidal Fermi surfaces in Fig. 4(a))[62–64]. This results in an about 10-times heavier confinement mass at the $\bar{\Gamma}$-point[62,64], which makes the quantum confinement weaker than at $\bar{M}$. At the $\bar{\Gamma}$-point, the splitting observed for Pb$_{0.93}$Ge$_{0.07}$Te is well-resolved at high momenta, but close to $\bar{\Gamma}$ there is a strong overlap of the subband dispersion. This is shown in the $\boldsymbol{k}\cdot\boldsymbol{p}$ calculations displayed on the right-hand side of Fig. 4(c). For this reason, the individual subbands are difficult to distinguish by ARPES. This strong overlap is absent for the $\overline{M}$-point (Fig. 4(e)) and thus, the Rashba spin-splitting induced by Ge doping is clearly resolved for each individual subband. Therefore, the direct comparison with the $\mathbf{k}\cdot\mathbf{p}$ calculations including the Rashba effect is facilitated and evidences a perfect agreement between theory and experiments. These results unequivocally demonstrate the induction of Rashba spin splitting by Ge doping and thus, that the $Pb_{1-x}Ge_xTe$ is a FERSC.

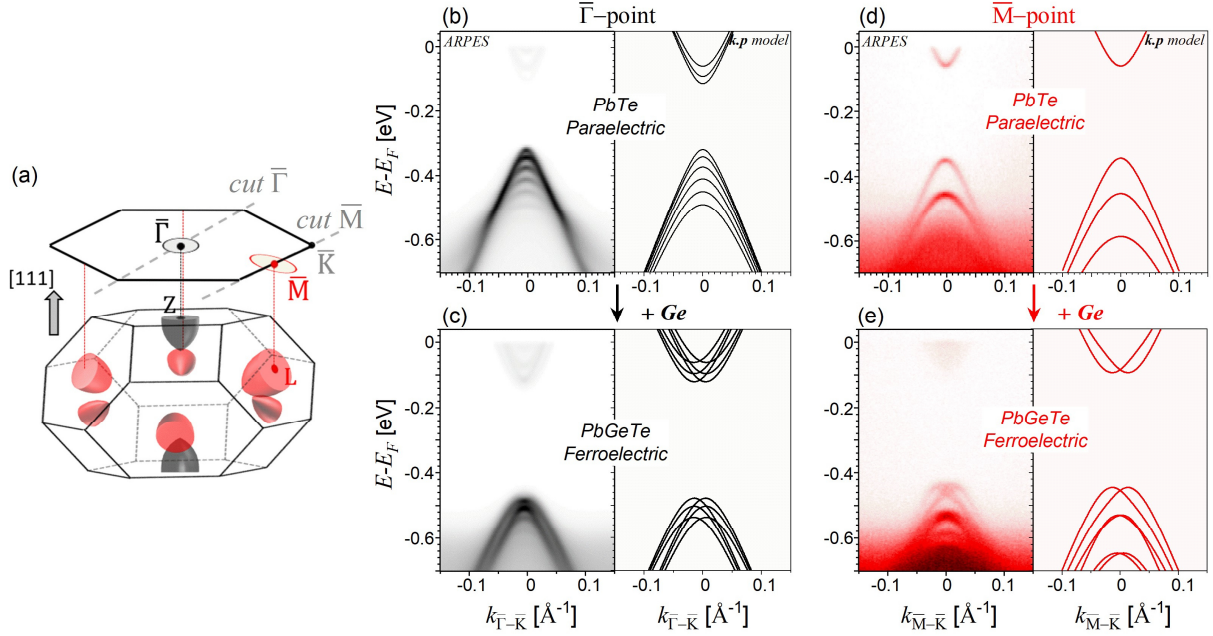

**Figure 4. Ferroelectricity and Rashba effect induced by Ge doping. (a)** 3D BZ of the $Pb_{1-x}Ge_xTe$ rhombohedral lattice with an elongation along [111] as well as the schematic Fermi pockets in red and black. The 2D projection of the BZ and of the Fermi pockets on the (111) surface is illustrated along with the ARPES measurement directions. **(b,c)** ARPES measurements at T=10 K along the $\overline{K\Gamma K}$ direction for the PbTe (b) and the $Pb_{0.93}Ge_{0.07}Te$ (c) 9 nm QWs. The fits using the $\mathbf{k}\cdot\mathbf{p}$ model develop in this work are plotted on the right of the corresponding ARPES spectra. **(d,e)** Similar than (b,c) along the $\overline{KMK}$ direction.

To further characterize the FERSC of $Pb_{1-x}Ge_xTe$, we have determined the temperature dependence of the Rashba splitting, focusing on the ARPES dispersions at the $\overline{M}$-points, where the quantized subbands are better resolved (see Fig. 4(d,e)). The results are shown in Fig. 5(a,b) where the ARPES maps recorded at 10<T<200 K are compared between the PbTe and the $Pb_{0.93}Ge_{0.07}Te$ cases. Evidently, no Rashba splitting is observed for the PbTe QW in the whole temperature range because it remains PE down to 10 K. As a result, the electronic subbands remain spin degenerate, meaning that neither a FE polarization nor any surface band bending is present in this sample. This is coherent with the fact that PbTe crystallizes in a cubic lattice and no rhombohedral distortion emerges at any positive temperature. Therefore, PbTe stands as a reference cubic phase persisting down to low temperature.

For the $Pb_{0.93}Ge_{0.07}Te$ QW, the Rashba spin-splitting observed at low temperature is found to gradually decreases with increasing temperature such that at T=200 K and above, only single sharp bands are observed. Thus, the Kramers degeneracy is lifted only below $T_c \sim 190$ K where the system is in the FE phase. As no external nor internal magnetism, nor an external electric field is present in our experiment, this is an unambiguous evidence for the ferroelectric Rashba effect. We can also safely rule out the presence of a surface-induced Rashba effect caused, e.g., by the presence of an accumulation or a depletion layer at the surface because this would result in a temperature independent Rashba spin-splitting and because it should appear also for the PbTe reference sample – which is clearly not observed. Therefore, the Rashba spin splitting emerging at low temperature is exclusively due to the intrinsic electric field appearing due to the FE polarization of the rhombohedral phase. The temperature dependent Rashba splitting is thus another key signature of the FE phase transition occurring in $Pb_{1-x}Ge_xTe$. Similar data was obtained for additional $Pb_{1-x}Ge_xTe$ samples with different Ge concentrations, as shown in the supplementary material.

The FE phase transition in $Pb_{1-x}Ge_xTe$ also induces an anomalous behavior of the energy gap versus temperature, as shown in Fig. 5(c). This gap was extracted from the ARPES spectra of the $Pb_{1-x}Ge_xTe$ QWs by measuring the energy separation between the electron and hole ground confined states at $k = 0$. We clearly see that the critical temperature $T_c$ coinciding with the emergence of the Rashba splitting, the slope of the gap versus temperature shows a clear sign change and thus, increases at low temperatures[53,65]. This is in complete contrast to the smooth temperature dependence measured for PbTe (Fig. 5(a)) monotonically decrease with decreasing temperature in perfect agreement with the literature[65,66]. The gap anomaly of $Pb_{1-x}Ge_xTe$ is thus another clear demonstration of the FE phase transition, its origin is discussed in the theory part detailed in the Methods section. Taken all together, our experimental data evidences the simultaneous occurrence of the FE phase transition, the Rashba effect and the gap anomaly in thin film $Pb_{1-x}Ge_xTe$ heterostructures at exactly the same critical temperature. Most importantly, repeated measurements show that the gap anomaly and the Rashba splitting emergence is reproducible and completely reversible with temperature.

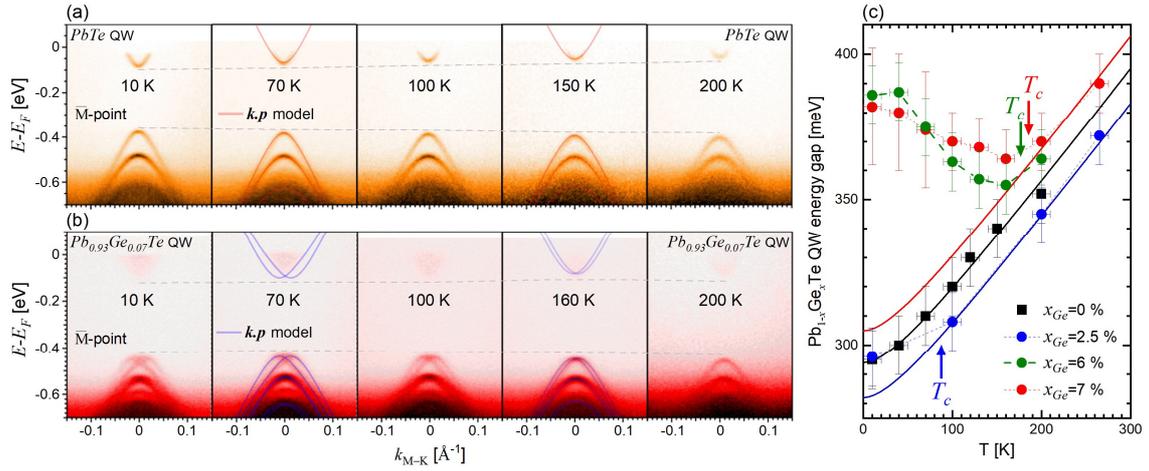

**Figure 5. Temperature dependence of the Rashba effect. (a,b)** ARPES spectra of PbTe (orange) and $Pb_{0.93}Ge_{0.07}Te$ (red) 9 nm-thick QWs at different temperatures from 10 to 200 K. The red and blue lines denote the fit using the $\boldsymbol{k} \cdot \boldsymbol{p}$ model shown for T=70 K, T=150 K and T=160 K. **(c)** QW band gaps versus temperature of the four investigated ARPES samples. The green (blue) data points correspond to a $Pb_{0.94}Ge_{0.06}Te$ ($Pb_{0.975}Ge_{0.025}Te$) QW shown in the supplementary material. The dashed lines are guide-for-the-eyes and the solid lines represent the expected gap dependence in the cubic phase varying like $0.5\,T^2/(T + 55)$ meV with temperature[65]. The critical temperatures are indicated.

In order to get more quantitative insight on the intrinsic Rashba effect and its interplay with the FE lattice distortion, we have modelled the ARPES data with a refined $\boldsymbol{k} \cdot \boldsymbol{p}$ model adapted from Ref.[51] and extended to describe the electronic structure of QWs in their rhombohedral phase. The modelling was done by decomposing the FE distortion into a strain effect and a sublattice shift of the anions/cations. The strains occurring during the rhombohedral distortion are taken into account by diagonal matrix elements in the $\boldsymbol{k} \cdot \boldsymbol{p}$ Hamiltonian that enlarge the energy gap. The sublattice shift responsible for the electric dipole is considered by adding two interband coupling parameters. They appear as k-linear terms in the $\boldsymbol{k} \cdot \boldsymbol{p}$ Hamiltonian, and account for the well-known Rashba parameters, $\alpha_R$, which governs the band splitting observed here in ARPES.

Calculations are described in the Methods section and with appropriate choice of parameters precisely agree with ARPES data, as shown by the solid lines Fig. 5 that represent the calculated QW band dispersion for selected temperatures of 50 and 150K. It is emphasized that the fit of the calculations to ARPES is excellent both for PbTe and Pb$_{1-x}$Ge$_x$Te in their respective cubic and rhombohedral phases. In particular, for the FE phase the calculations involve the additional k-linear terms (or Rashba term) to describe the observed Rashba spin splitting, while these are absent in the cubic phase. The Rashba parameter $\alpha_R$ obtained by the fit of the experiments are shown in Fig. 6(a) versus temperature for three samples with different $x_{Ge}$. In all cases, the Rashba parameter is found to increase when temperature is decreased and that it is zero for $T > T_C$. This parameter, which is responsible for the spin-splitting of the bands, is directly related to the FE polarization via the optical deformation potential $\Xi_o$[51]. Indeed, it writes:

$$\alpha_R = \frac{\hbar v \Xi_o \sqrt{3}}{E_g}\delta = \frac{\hbar v \Xi_o}{2aqNE_g}P_{elec} \quad (1)$$

Where $E_g$ is the energy gap of the QW material, $v$ is the Dirac velocity and $\delta$ is the off-center lattice shift in units of the lattice parameter ($\delta u = 2a\sqrt{3}\delta$) as already defined in the previous section. The polarization is written as $P_{elec} = Nq\delta u = 2aqN\sqrt{3}\delta$, where $q$ is the charge of the dipole moment and $N$ the number of dipoles per volume unit. The main result here is that the Rashba parameter is directly proportional to the sublattice shift, thus, the FE polarization.

In the Landau-Ginzburg theory, the polarization stands as the order parameter defining the FE phase transition. Within the framework of this theory, one gets $P_{elec} \propto \sqrt{T_C - T}$ in the FE phase[67,68] and thus, $\alpha_R \propto \sqrt{T_C - T}$ following Eq. (1), if one neglects the small temperature variation of $E_g$ and $a$. This behavior is fully verified by the perfect fit of $\alpha_R \propto \sqrt{T_C - T}$ to the experimental data as predicted by the Landau-Ginzburg theory (see the solid lines in Fig. 6(a)). The fit allows an accurate determination of the coefficient $C$, which is plotted versus the Ge in Fig. 6(b) and is phenomenologically well-described by $C(x_{Ge}) = 0.57\sqrt{x_{Ge}}$ in unit of eV.Å.K$^{-1/2}$ within the investigated composition range $0 < x_{Ge} < 0.07$. These results imply that the FE phase transition in our Pb$_{1-x}$Ge$_x$Te QW heterostructures is very well-described by a second order phase transition. At low temperature, the Rashba constant reaches 2 eV.Å in the Pb$_{0.93}$Ge$_{0.07}$Te QW heterostructure. It is comparable to state-of-the-art Rashba systems like BiTeI ($\alpha_R = 3.8$ eV.Å)[34], SnTe ($\alpha_R = 4.4$ eV.Å)[6], Bi-doped Pb$_{1-x}$Sn$_x$Te ($\alpha_R = 2 - 4$ eV.Å)[43] or even GeTe ($\alpha_R = 4.2 - 4.8$ eV.Å)[39], indicating a giant Rashba effect in Pb$_{1-x}$Ge$_x$Te.

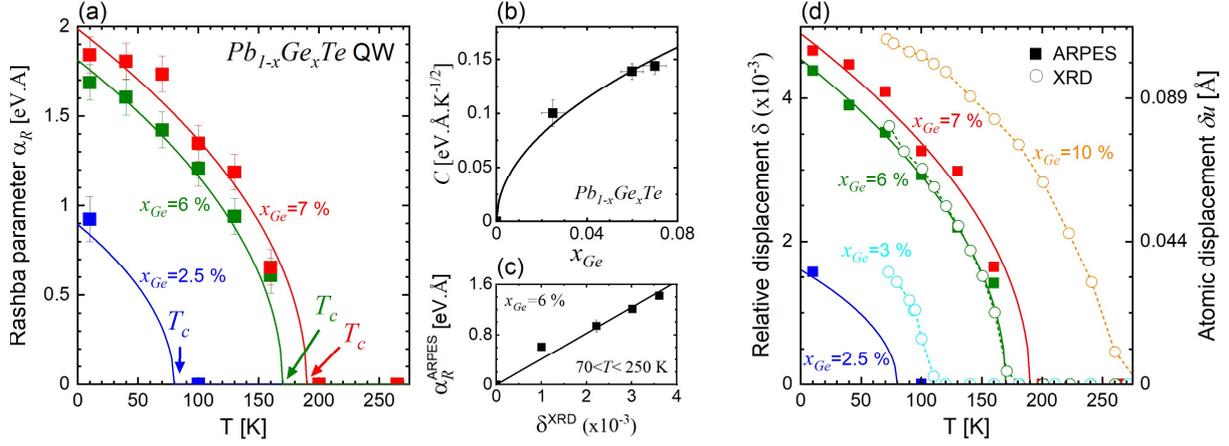

**Figure 6. Interplay between structural and electronic properties of a FERSC. (a)** Rashba parameters of $Pb_{1-x}Ge_xTe$ versus temperature determined by ARPES. The solid lines correspond to the $\alpha_R(T) = C\sqrt{T_C - T}$ dependence expected from the Landau-Ginzburg theory for a second-order phase transition. **(b)** Plot of the coefficient $C$ determined in (a) versus the Ge concentration (dots). The solid line shows the experimental fit using $C(x_{Ge}) = 0.57\sqrt{x_{Ge}}$. **(c)** Rashba parameters determined in ARPES between 70 and 250 K for the $Pb_{0.94}Ge_{0.06}Te$ QW (green dots in (a)) as a function of the relative atomic displacement measured in XRD (see Fig. 3(f)). The solid line is computed using Eq. (1) with $\Xi_o = 16.5$ eV. **(d)** Atomic relative displacement $\delta$ as a function of temperature. The square dots are the value deduced from the Rashba constant determined by ARPES, and the circles represent the values measured in XRD (see Fig. 3(f)). Dashed lines are guide-for-the-eyes. Solid lines represent the $\sqrt{T_C - T}$ dependence deduced from (a) using Eq. (1) with $\Xi_o = 16.5$ eV. The sublattice shift $\delta u = 2\delta\sqrt{3}a$ that is shown on the right axis is calculated using a temperature constant lattice parameter $a = 6.425$ Å.

Most importantly, based on our XRD and ARPES data, we can conclude for the first time on the quantitative relation between the FE polarization – the sublattice shift – and the electronic spin texture – the Rashba spin splitting – in a FERSC. This is shown by Fig. 6(c), where the Rashba parameter $\alpha_R$ determined in ARPES is plotted as a function of the relative atomic displacement $\delta$ measured by XRD for $Pb_{0.94}Ge_{0.06}Te$. The linear relation between these two quantities (see Eq. (1)) is fully confirmed and yields the optical deformation potential $\Xi_o = 16.5$ eV, in good agreement with literature[51,53].

In order to illustrate the unique structure-electronic correlation of FERSC, Fig. 6(d) shows the temperature evolution of $\delta$ measured by XRD, as well as the $\delta$ deduced from the Rashba constant $\alpha_R$ measured by ARPES using Eq. (1). The best fit between ARPES and XRD data directly yields the optical deformation potential $\Xi_o = 16.5$ eV for the all range of investigated $Pb_{1-x}Ge_xTe$. This directly demonstrates that the sublattice shift in a FERSC is the key source responsible for the Rashba splitting, and that the displacement strength controls the Rashba constant. These results show the equivalence between structural and electronic properties, i.e., between the FE polarization strength and Rashba spin splitting, in FERSC.

## 3. Conclusion

In summary, we have described the MBE growth and characterized the structural and electronic properties of epitaxial $Pb_{1-x}Ge_xTe$ single crystalline layers over a large range of temperature and Ge contents, qualifying this material as a FERSC with outstanding properties sustained even in the thin film limit. The FE structural phase transition was revealed by temperature dependent XRD experiments, showing that the sublattice shift responsible for the non-centrosymmetry reaches values exceeding 0.1 Å already for $x_{Ge} < 10$ %. We demonstrate that the FE phase transition leads to a giant Rashba effect observed by ARPES, as large as 2 eV.Å. The temperature dependent Rashba constant precisely follows the behavior depicted by the Landau-Ginzburg theory of a second-order phase transition. Furthermore, the magnitude of the Rashba effect is linearly related to the anion versus cation atomic displacement, and thus, to the electric polarization.

In this way, $Pb_{1-x}Ge_xTe$ stands for a highly promising FERSC system because it features a number of advantages over other FERSC materials, namely, (i) a low doping level, (ii) a direct optical gap, (iii) a tunable and high critical temperature due to the ternary nature of the $Pb_{1-x}Ge_xTe$ alloy, and (iv) a high Rashba spin-splitting in nanometric layers. This opens up new avenues for the realization of FERSC devices dedicated to a large number of applications.

*Methods*

**Growth.** MBE growth of Pb$_{1-x}$Ge$_x$Te layers and QWs on BaF$_2$ (111) substrates was performed using a VARIAN Gen II MBE system under ultra-high vacuum (UHV) conditions (2 x 10$^{-10}$ mbar) using PbTe, GeTe and Bi$_2$Te$_3$ as source materials. The composition of the ternary layers Pb$_{1-x}$Ge$_x$Te was controlled by the GeTe/PbTe beam flux ratio measured precisely using a quartz microbalance moved into the substrate position and the sample temperature measured with an infrared pyrometer. The growth was monitored *in-situ* using RHEED and the sample surface characterized by AFM using a Veeco Dimensions 3100 SPM. Bulk-like films with 4 – 5 µm thickness with different composition up to x$_{Ge}$ = 0.13 were grown for the temperature dependent XRD studies to assess the FE phase transitions. For ARPES, PbTe and Pb$_{1-x}$Ge$_x$Te QW films of 8-10 nm thickness were grown on 100 nm wide band gap Pb$_{0.9}$Eu$_{0.1}$Te barrier layers in order to achieve a quantum confinement of the electronic states in the Pb$_{1-x}$Ge$_x$Te layers. These samples were intentionally highly n-doped using in-situ Bi doping in order to bring the Fermi level high into the conduction band to allow the observation of both hole and electron states by ARPES.

**Structure characterization and FE phase transition.** The sample structure was characterized in detail to determine the lattice parameter, thickness and layer composition using a standard Pananalytical materials research diffractometer as well as a Rigaku SmartLab x-ray diffractometer equipped with a custom-made variable temperature sample stage with hemispherical PEEK window (ColdEdge International) for temperature dependent measurements down to 72 K using pumped liquid nitrogen. Both instruments are equipped with a Cu x-ray source and a Ge(220) channel-cut monochromator. The in-plane and out-of-plane lattice parameters as well as the FE lattice distortion was determined from reciprocal space maps recorded around the (333), (444) and (xxx) reflections.

**Angle-resolved photoemission spectroscopy.** ARPES measurements were performed at the SOLARIS synchrotron at the high-resolution URANOS beamline at the SOLARIS synchrotron in Krakow, Poland. For this purpose, the samples were transferred from the MBE to the synchrotron under UHV conditions using a battery-operated Ferrovac vacuum suitcase. UV radiation of 18 eV was used for excitation of the photoelectrons and their angular and energy distribution measured by a VG Scienta DA30L electron spectrometer with energy and angular resolution better than 3 meV and 0.1°, respectively. Temperature-dependent measurements were taken under ultra-high vacuum condition (<10$^{-10}$ mbar) with a horizontally polarized light and a vertical slit.

**Modelling of the band structure in FERSC heterostructures using $k.p$ theory.** The $k.p$ Hamiltonian to model the PbGeTe QW heterostructures is modified from the Hamiltonian given in Ref.[51] for rhombohedral bulk PbGeTe by including the z-dependent quantum confinement potential. It writes:

$$H = \begin{pmatrix} -V(z) + \alpha_{2,\parallel}k_x & -i\alpha_{1,\parallel}k_- - i\alpha_{2,z}\dfrac{d}{dz} & -i\hbar v_z \dfrac{d}{dz} & (u^2 - v_1^2)v_\parallel \hbar k_- \\ i\alpha_{1,\parallel}k_+ - i\alpha_{2,z}\dfrac{d}{dz} & -V(z) - \alpha_{2,\parallel}k_x & (u^2 - v_1^2)v_\parallel \hbar k_+ & i\hbar v_z \dfrac{d}{dz} \\ -i\hbar v_z \dfrac{d}{dz} & (u^2 - v_1^2)v_\parallel \hbar k_- & V(z) + \alpha_{2,\parallel}k_x & -i\alpha_{1,\parallel}k_- - i\alpha_{2,z}\dfrac{d}{dz} \\ (u^2 - v_1^2)v_\parallel \hbar k_+ & i\hbar v_z \dfrac{d}{dz} & i\alpha_{1,\parallel}k_+ - i\alpha_{2,z}\dfrac{d}{dz} & V(z) - \alpha_{2,\parallel}k_x \end{pmatrix}$$

Here, $v_z$ and $v_\parallel$ are the out-of-plane and in-plane Dirac velocities accounting for the anisotropy; $k_\pm = k_x \pm ik_y$; $V(z)$ is the confinement potential that is $E_g/2$ in the PbGeTe quantum well, 450 meV in the Pb$_{0.9}$Eu$_{0.1}$Te barrier underneath, and is taken as 2 eV to mimic the vacuum barrier at the surface. The terms in red appear with the FE distortion. We have set:

$$\alpha_{1,2,\parallel,z} = 2uv_{1,2}\hbar v_{\parallel,z}$$

where $u$ and $v_{1,2}$, introduced in Ref.[51], parametrize the lattice distortion. Indeed, $u = 1$ denotes the cubic phase (with $v_1 = 0$ and $v_2 = 0$), and the ferroelectric polarization increases as $u$ moves away from unity to lower values. They are given by:

$$\begin{cases} u = \cos\theta \\ v_1 = \sin\theta \cos\phi \\ v_2 = \sin\theta \sin\phi \end{cases} \quad (2)$$

So that $u^2 + v_1^2 + v_2^2 = 1$. Here, the angles $\theta$ and $\phi$ are defined in terms of matrix elements $\Delta_1$ and $\Delta_2$, which couple orbitals and orbitals and spins, respectively.

$$\begin{cases} \tan\phi = \dfrac{\Delta_2}{\Delta_1} \\ \tan 2\theta = \dfrac{1}{\sqrt{\dfrac{E_g^2}{4(\Delta_1^2 + \Delta_2^2)} - 1}} \end{cases} \quad (3)$$

Additional matrix elements have been included to take into account the dilatation and shear strains emerging in the rhombohedral phase $\gamma = \delta_1 - \delta_2$ following the notation used in Ref.[51]. They emerge as diagonal terms in the Hamiltonian and are taken into account in the energy gap parameter $E_g$. Consequently, the PbGeTe energy gap is renormalized to:

$$E_g = \sqrt{(2\Delta + \gamma)^2 + 4(\Delta_1^2 + \Delta_2^2)} \quad (4)$$

where $2\Delta$ is the energy gap if the lattice was in the cubic phase. This explains the anomalous temperature dependence of the energy gap observed in Fig. 5(c). However, the lack of independent information on $\gamma$ and $\Delta_2$ (see below) prevents us from drawing any further conclusions.

In summary, the rhombohedral distortion introduces the four following parameters. For further details, the reader is referred to Ref.[51], and to equivalent theories developed in Ref.[53,69].

- Intraband matrix elements $\delta_{1,2}$ that account for the strains.
- Interband matrix elements $\Delta_{1,2}$ accounting for the electric dipole, emerging from the sublattice shift.

To find the confined state dispersions of the FERSC quantum well, we first solve exactly the following Hamiltonian:

$$H_0 = \begin{pmatrix} -V(z) & 0 & -i\hbar v_z \dfrac{d}{dz} & 0 \\ 0 & -V(z) & 0 & i\hbar v_z \dfrac{d}{dz} \\ -i\hbar v_z \dfrac{d}{dz} & 0 & V(z) & 0 \\ 0 & i\hbar v_z \dfrac{d}{dz} & 0 & V(z) \end{pmatrix}$$

This has been performed in Ref.[70,71]. One obtains the energy of the confined states as well as their wavefunctions $\psi_n$. The additional terms in $k_x$, $k_y$ and $\alpha_{2,z}$ are taken into account in a perturbation theory. The matrix $\langle \psi_n | \delta H | \psi_m \rangle$ with $n$ and $m$ denoting the confined states, and $\delta H = H - H_0$ is then solved numerically and gives the k-dispersions of the confined states. Moreover, for a thin layer, the ferroelectric distortion is likely to occur in the direction perpendicular to the surface, and thus, leads to one single domain. This is demonstrated by the XRD measurements performed on thick films and detailed in Fig. 3(b,c), showing that the prominent domain is the one with the elongation perpendicular to the surface. In this way, one gets $\Delta_2 = 0$ at the $\bar{\Gamma}$-point by symmetry consideration[51][69]. More generally, the matrix elements $\gamma$, $\Delta_1$ and $\Delta_2$ have no reason to be equal at the $\bar{\Gamma}$ and the $\bar{M}$-points.

*At the $\bar{\Gamma}$-point*, the results are analytical under a good approximation. The $n^{th}$ subband dispersions at the $\bar{\Gamma}$-point for Pb$_{1-x}$Ge$_x$Te in its rhombohedral phase write:

$$E_n = \pm\sqrt{\Delta_n^2 + \hbar^2 v^2 k^2 \pm 2\alpha_R \Delta_n |k|} \tag{5}$$

Here, $\Delta_n$ is the energy of the $n^{th}$ subband at $k = 0$ obtained by solving $H_0$, and $\alpha_R$ the Rashba constant. Equation (5) stands as the dispersion of Rashba-split Dirac bands. Note that the dispersion is isotropic at the $\bar{\Gamma}$-point. If $E_R$ is defined as usual as the energy difference between the top of the valence band and the energy at $k = 0$; and $k_R$ as the position of the valence band maximum[4,72], then the Rashba parameter $\alpha_R$ writes:

$$E_R = \Delta_n\left[1 - \sqrt{1 - \frac{\alpha_R k_R}{\Delta_n}}\right] \Leftrightarrow \alpha_R = \frac{2E_R}{k_R}\left[1 - \frac{E_R}{2\Delta_n}\right] \tag{6}$$

This formula has been derived in the framework of a 2-band Dirac model, thus, well-adapted for narrow gap materials. For relatively wide gap materials, Eq. (6) is well approximated by the formula obtained with a one band parabolic model $\alpha_R = 2E_R/k_R$ [4,72]. The factor $1 - E_R/(2\Delta_n)$ stems from the Dirac nature of the material. Note that at high momenta, Eq. (5) for the valence states tends towards:

$$E_n = -\hbar v k \pm \frac{\alpha_R \Delta_n}{\hbar v}$$

Thus, at a fixed high momentum, the spins are shifted by $2\alpha_R \Delta_n/\hbar v$ in energy, which gives an experimental estimation of the Rashba constant $\alpha_R$ at the $\bar{\Gamma}$-point. For the Pb$_{0.93}$Ge$_{0.07}$Te QW at low temperature (see Fig. 4(c)), one finds $\alpha_R \sim 0.8$ eV.Å. This gives a relatively small value of $\Delta_1 \sim 12$ meV and thus, a non-negligible $\gamma$ value to explain the gap anomaly observed in Fig. 5(c) and in the supplementary material (see Eq. (4)).

*At the $\bar{M}$-points*, one needs to rotate $H$ to align the $z$ axis with the great axis of the oblique valley, which is tilted by 70.5 °. Such a rotation is described in Ref.[71] and prevents us from using analytical expressions. Moreover, the ARPES measurements are taken along the $\bar{M} - \bar{K}$ direction (see Fig. 4(a)), which corresponds to $k_y$ in our coordinates. Therefore, the dispersions are calculated for $k_x = 0$, which means that the parameter $\Delta_2$ (or $v_2$) does not intervene in the fit of the measured dispersions apart from its influence on the energy gap (see Eq. (4)). We chose to take $E_g$ as the fitting parameter, and to arbitrary put $\Delta_2 = 0$. In this way, the fitting parameters are $E_g$, $v_z$, $v_\parallel$ and $\Delta_1$ (or $u$, or $v_1$ as $v_1 = \sqrt{1-u^2}$). They are listed in the supplementary material. The parameter $\Delta_1$ is accurately deduced as it is the only parameter responsible for the well-resolved band spin-splitting. For $\Delta_2 = 0$, Eq. (2) and (3) give $uv_1 E_g = \Delta_1$ so that the Rashba parameter $\alpha_R$ introduced above writes:

$$\alpha_R = \alpha_{1,\parallel} = 2uv_1\hbar v_\parallel = 4\hbar v_\parallel \frac{\Delta_1}{E_g}$$

$\Delta_1$ has the form of an interband optical deformation potential $\Xi_o$ and writes $\Xi_o\sqrt{3}\,\delta/4$. Thus, one retrieves Eq. (1).

*Acknowledgements*

We acknowledge support by the Austrian Science Fund (FWF, Project I-4493). The ARPES setup was developed under the provision of the Polish Ministry and Higher Education project Support for research and development with the use of research infrastructure of the National Synchrotron Radiation Centre "SOLARIS" under contract No 1/SOL/2021/2. CzechNanoLab project LM2023051 funded by MEYS CR is gratefully acknowledged for the financial support of the measurements at CEITEC Nano Research Infrastructure. The authors acknowledge the funding from the project CZ.02.01.01/00/22_008/0004572 "Quantum materials for applications in sustainable technologies".

*Conflict of Interest*

The authors declare no conflict of interest.

*Data availability statement*

All the data supporting the finding of this study are available from the corresponding authors upon reasonable request.